\documentclass[doublecol]{epl2} 

 \usepackage{amssymb}

\usepackage{color}

\title{Phase order in superfluid helium films}
\shorttitle{Phase order in superfluid helium films} 

\author{Steven T. Bramwell,\inst{1}\footnote{email: s.t.bramwell@ucl.ac.uk} Michael F. Faulkner,\inst{1,2} Peter C. W. Holdsworth\inst{2}
\and Andrea Taroni\inst3\footnote{Present address: Nature Physics, 4 Crinan Street, London N1 9XW, UK.}}
\shortauthor{Bramwell \etal}

\institute{                    
\inst{1}London Centre for Nanotechnology and Department of Physics and Astronomy,
University College London, 17-19 Gordon Street, London WC1H 0AJ, United Kingdom.
\inst{2}Laboratoire de Physique, Universit\'{e} de Lyon, \'{E}cole Normale Sup\'{e}rieure de Lyon,
46 all\'{e}e d'Italie, 69364 Lyon Cedex 07, France.
\inst{3}Department of Physics and Astronomy, Uppsala University, Box 516, 751 20 Uppsala, Sweden.
}

\abstract{
Classic experimental data on helium films are transformed to estimate a finite-size phase order parameter that measures the thermal degradation of the condensate fraction in the two-dimensional superfluid. The order parameter is found to evolve thermally with the exponent  $\beta = 3 \pi^2/128$, a characteristic, in analogous magnetic systems, of the Berezinskii-Kosterlitz-Thouless (BKT) phase transition. Universal scaling near the BKT fixed point generates a collapse of experimental data on helium and  ferromagnetic films, and implies new experiments and theoretical protocols to explore the phase order. These results give a striking example of experimental finite-size scaling in a critical system that is broadly relevant to two-dimensional Bose fluids.  }

\begin{document}

\maketitle

\section{Introduction}

The remarkable properties of  liquid helium II -- such as its ability to creep over the walls of its container -- establish it as arguably the most interesting state of condensed matter~\cite{Atkins,Donelly}. 
It may be represented as a superposition of two fluids: a normal fluid and a superfluid with zero viscosity and zero entropy~\cite{Tisza}. To deduce the superfluid fraction, Andronikashvili~\cite{And} utilised the fact that only the normal fluid, with its finite viscosity, moves with a stack of closely spaced discs. By measuring the frequency of torsional oscillation, the normal density $\rho_{\rm n}$, and hence the superfluid density, $\rho_{\rm  s} = \rho-\rho_{\rm  n}$, could be measured (here $\rho$ is the total density). The superfluid fraction becomes finite at the lambda point (2.2 K), increasing towards unity at zero temperature, where helium behaves as an ideal Eulerian fluid with only irrotational flow~\cite{Donelly}. Many years later, Bishop and Reppy used an adaption of Andronikashvili's method with an oscillating substrate to measure the temperature evolution of $\rho_{\rm  s}$ for ultrathin helium films~\cite{BR}. 

Superfluidity is driven by Bose condensation and two-fluid hydrodynamics is a consequence of this~\cite{LL,Balibar}. The condensate wavefunction~\cite{PO}
is represented by a two-component field  $\psi({\bf r},t) = \sqrt{n_0} e^{i \Phi({\bf r},t)}$ (where $n_0$ is the helium number density), which can be considered to be classical in the region of the transition. The superfluid velocity is equal to the condensate velocity as determined by the phase of the condensate wavefunction, ${\bf v}_{\rm s} = (\hbar/m)\nabla \Phi$,  but the condensed fraction $f_{\rm c} = \langle \psi\rangle^2/n_0$ is only indirectly related to the superfluid fraction $f_{\rm s} = \rho_{\rm s}/\rho$. The nonideality of the Bose fluid, arising largely from the mutual repulsion of helium atoms, causes the condensate fraction to be much smaller than the superfluid fraction in the low-temperature limit: $f_{\rm c} \sim 0.075$, compared to $f_{\rm s}=1$~\cite{Glyde,Moroni}.  

Both superfluid and condensate fractions are degraded by thermal excitations. In bulk helium they fall to zero  in proportion at the lambda transition~\cite{Fisher}. In contrast, helium films have a very different response to thermal fluctuations. High- and low-temperature phases are separated by a vortex deconfinement transition in the velocity field, of the  Berezinskii-Kosterlitz-Thouless (BKT)~\cite{B,KT,K,JKKN} type.
While the superfluid density remains intensive below the BKT transition, the condensed fraction is non-intensive, decaying to zero in the thermodynamic limit at all finite temperatures~\cite{LL}. Formally,  two-dimensional helium II is a superfluid but not a condensate. 

The difference between two- and three-dimensional helium II arises because the low-energy spectrum of excitations above the superfluid ground state consists of gapless phonons~\cite{LL}. This places the two-dimensional system at its lower critical dimension, ensuring  critical correlations and leading to a formal absence of long-range order, in accord with the theorems of Hohenberg~\cite{Hohenberg} and Mermin-Wagner~\cite{MW}. However, the critical system is topologically ordered, which allows phonon excitations and superfluidity but excludes long-range phase coherence, as measured by a finite condensate fraction~\cite{NK}. 

The Bose fluid may be mapped in detail to the quantum S = 1/2 -XXZ magnetic model~\cite{MM}, a relation that has added considerably to the understanding of quantum magnets~\cite{Ruegg}. However, to 
 treat the BKT transition in helium films, it is sufficient to consider a classical 2D-XY model within the Villain approximation~\cite{NK}. In the critical region, an effective Hamiltonian for the superfluid fraction is 
\begin{equation} \label{eq1}
H_{\rm s} = (1/2) \Upsilon \int |\nabla \Phi({\bf r})|^2 d {\bf r},  
\end{equation}
where $\Upsilon = (\hbar/m)^2 \rho_{\rm s}$ is the helicity modulus, a measure of the phase stiffness~\cite{Fisher}. This maps to the Hamiltonian of a continuum classical harmonic spin wave model with angular variable $\Phi$ and effective exchange constant $J_{\rm eff} \mapsto \Upsilon$. The effective spin stiffness is then $K_{\rm eff} = J_{\rm eff}/kT \mapsto \Upsilon/kT$. Nelson and Kosterlitz~\cite{NK} calculated $K_{\rm eff}$ and hence $\rho_{\rm s}$ by integrating out thermally excited vortex pairs and absorbing their effect into a renormalised spin stiffness. In the critical regime the spin correlations decrease as $g(r) \sim r^{-\eta}$ with $\eta(T) = 1/2\pi K_{\rm eff}(T)$, an increasing function of temperature. At the BKT transition $\eta = 1/4$ and $K_{\rm eff}$ reaches the universal value $2/\pi$ before jumping discontinuously to zero. 

The Bishop--Reppy measurements of $\rho_{\rm s}(T)$ for  helium films gave very convincing support for the BKT transition~\cite{BR}.  They were analysed using the theory of Ambegaokar, Halperin, Nelson and Siggia (AHNS)~\cite{AHNS}, which accounted, using the renormalisation group theory~\cite{K,JKKN,NK}, for the linear dynamical response of both bound and free vortices, and their effect on spin waves or phase fluctuations. As the renormalisation group procedure relates the properties of systems of different sizes, it allows the calculation of scaling behaviour in the limit of large system size.  The experimental quantities analysed in this way were period shift and dissipation: the static $\rho_{\rm s}$ and hence $K_{\rm eff}$ or $\Upsilon$ can be related to a combination of these~\cite{BR}.

\section{Finite-size order}

A practical consequence of the criticality of the low-temperature phase and the slow decay of correlations with distance is that the Mermin-Wagner theorem, although valid, is circumvented in all experimental systems~\cite{BH,BH2}. This arises because the temperature-dependent exponent, $\eta(T)/2\le 1/8$, characterising the decay of order parameter correlations with distance, is always a number much less than unity in the low-temperature phase. The consequence is that even perfect experimental realisations of a 2D-XY system of size $L$ must show a finite order parameter of order $M(L) \sim \sqrt{g(L)} = L^{-\eta/2}$ that is still far from negligible even for macroscopic scales of experimental relevance. Helium films are no exception to this rule and a corresponding power law tail of the correlation function in reciprocal space has been observed by neutron scattering~\cite{Diallo}. Similarly, interest in the nature of the `quasi-condensate' in the two-dimensional Bose fluid has been generated by elegant new experiments on trapped atomic gases~\cite{HadNat,Had}.  This finite-size order is particularly interesting as it affords experimental access to finite-size scaling at a critical point. In this paper we explore the relation of superfluid density to finite-size order, as measured by the condensate fraction of helium films.

We approximate the helium film to a lattice 2D-XY model in Villain's approximation~\cite{Villain} and define the normalised  phase order parameter ampltitude $\Psi= \langle \cos(\Phi-\bar{\Phi}) \rangle $ where $\bar{\Phi}$ denotes an instantaneous average. This corresponds, for the helium film, to the magnetic order parameter of a thin-film magnet: $M \mapsto\Psi$. The square of the finite-size order parameter is a classical approximation to the thermal component of the condensate fraction, measured relative to its zero-temperature value, $\Psi^2 \sim (f_{\rm c}(T) -f_{\rm c}(0))/f_{\rm c}(0)$. The classical approximation breaks down at low temperatures, owing to quantisation of the normal modes~\cite{LL}.

The order parameters of 2D-XY models of finite extent have been studied in detail and shown to have several interesting properties. 
The instantaneous, spatially averaged, measure of the order parameter  is a vector with both phase and amplitude. Its phase diffuses slowly around a circle, while the distribution of its amplitude is surprisingly sharply peaked around the mean value~\cite{Archambault,PRE}, corresponding to a `Mexican hat' potential, even along the critical line. At the BKT transition, the temperature evolution of the amplitude, $\Psi$, mimics the power law behaviour of a conventional ferromagnetic transition, with effective critical exponent $\beta = 3 \pi^2/128 \simeq 0.23$ (Ref.~\cite{BH,BH2}) -- one could say that the quasi-long range order is formed at a quasi-ferromagnetic transition. The universal exponent was calculated in the scaling limit by the renormalisation group method and confirmed in finite-size systems by numerical simulations. 

It is worth emphasising that while this $\beta$ is not a conventional critical exponent, it is nevertheless a striking signature of the BKT transition that has been widely reported in experiments. These include experiments on layered magnets~\cite{Als,Greven,Melzi,Blundell}, ultra thin magnetic films~\cite{Willis, Elmers, Hjorvarsson} and two-dimensional melting~\cite{Nuttall}. It has been equally evident in quasi-classical magnets~\cite{Als} and in quantum magnets~\cite{Greven,Melzi,Blundell}, as measured by neutron scattering~\cite{Als}, $\mu$SR~\cite{Blundell}, magneto-optic Kerr effect~\cite{Willis,Hjorvarsson} and spin-polarized electron diffraction~\cite{Elmers}.  Ref. \cite{Andrea} discusses the range of applicability of this result, the distinction of this $\beta$ from conventional critical exponents associated with symmetry-breaking fields,  and a review of the experimental literature up until 2008. More recent observations include, for example, Refs.~\cite{Kraemer, Vale}.
The effective exponent $\beta = 3 \pi^2/128$ has also been recommended as a diagnostic for the BKT transition in numerical studies~\cite{Wexler}, and has been previously discussed in relation to models of two-dimensional Bose-Einstein condensates~\cite{Trom}.  

The question naturally arises: do helium films show the same scaling of the finite-size order parameter as that exhibited by 2D-XY magnets? A direct test of the question is possible in principle~\cite{Diallo}.
Here we propose an experimental protocol that estimates  the phase order parameter $\Psi(L)$ indirectly from the superfluid density measured by Bishop and Reppy~\cite{BR}. In analogy to magnetic systems, we relate $\Psi$ to $\Upsilon$ by the following equation, 
\begin{equation}\label{BH}
\Psi(L,T) \simeq \left(\frac{1}{\gamma L^2} \right)^{kT/8\pi \Upsilon(L,T)} 
\end{equation} 
where $L$ is a dimensionless measure of the system size and $\gamma = 1.8456\dots$~\cite{PRE}.  The above equation comes from equating the finite-size helicity modulus divided by temperature, $\Upsilon(L,T)/kT$ with the scale ($r$) dependent  spin stiffness $K_{\rm eff}(r)|_{r = L}$, as defined in Refs. \cite{JKKN,NK}, and approximating the 
finite-size order parameter to  $\Psi(L,T) \simeq (\gamma L^2)^{-1/8 \pi K_{\rm eff}(L,T)}$, as in Ref. \cite{BH}. 

A necessary condition to obtain Eq. (\ref{BH}) is that $K_{\rm eff}(r,T)$ is approximately constant over length scales of order $L$, which is the case for $K_{\rm eff}(L,T)\simeq 2/\pi$. Away from this universal point, corrections arise from the $r$-dependence of $K_{\rm eff}(r)$~\cite{Archambault}. The theory of Ref.~\cite{BH} that yields $\beta = 3\pi^2/128$, like that of AHNS, uses the nonlinear renormalisation group equations of Nelson and Kosterlitz \cite{NK}. However, taking $K_{\rm eff}(r)$ to be constant over scales of order $L$ enforces a neglect of dissipation in the AHNS dynamical treatment.
  
Eq. (\ref{BH}) is an experimentally realisable example of critical finite-size scaling, which here is accessible because of the continuous line of critical points below the BKT transition temperature. Dimensional homogeneity requires that scaling behaviour is confined to an `inertial range'~\cite{BramwellPRL} between two length scales~\cite{Goldenfeld,Barenblatt}. The first is the microscopic length $a$, which for helium is the inter-particle spacing and for the 2D-XY model is the lattice constant. The second is a much larger length scale $l$, imposed by experimental conditions such as an anisotropy gap, a disorder-persistence length, a dynamical scale (as in the case analysed below), or ultimately the system boundaries.  The integral scale $L= l/a$ is then a dimensionless group that enables the order parameter to acquire the anomalous dimension $ kT/8 \pi \Upsilon(L,T)$, as demonstrated by  Eq. (\ref{BH})~\cite{PRE}.


\begin{figure}[h!]
\begin{center}
\onefigure[width=0.75\linewidth]{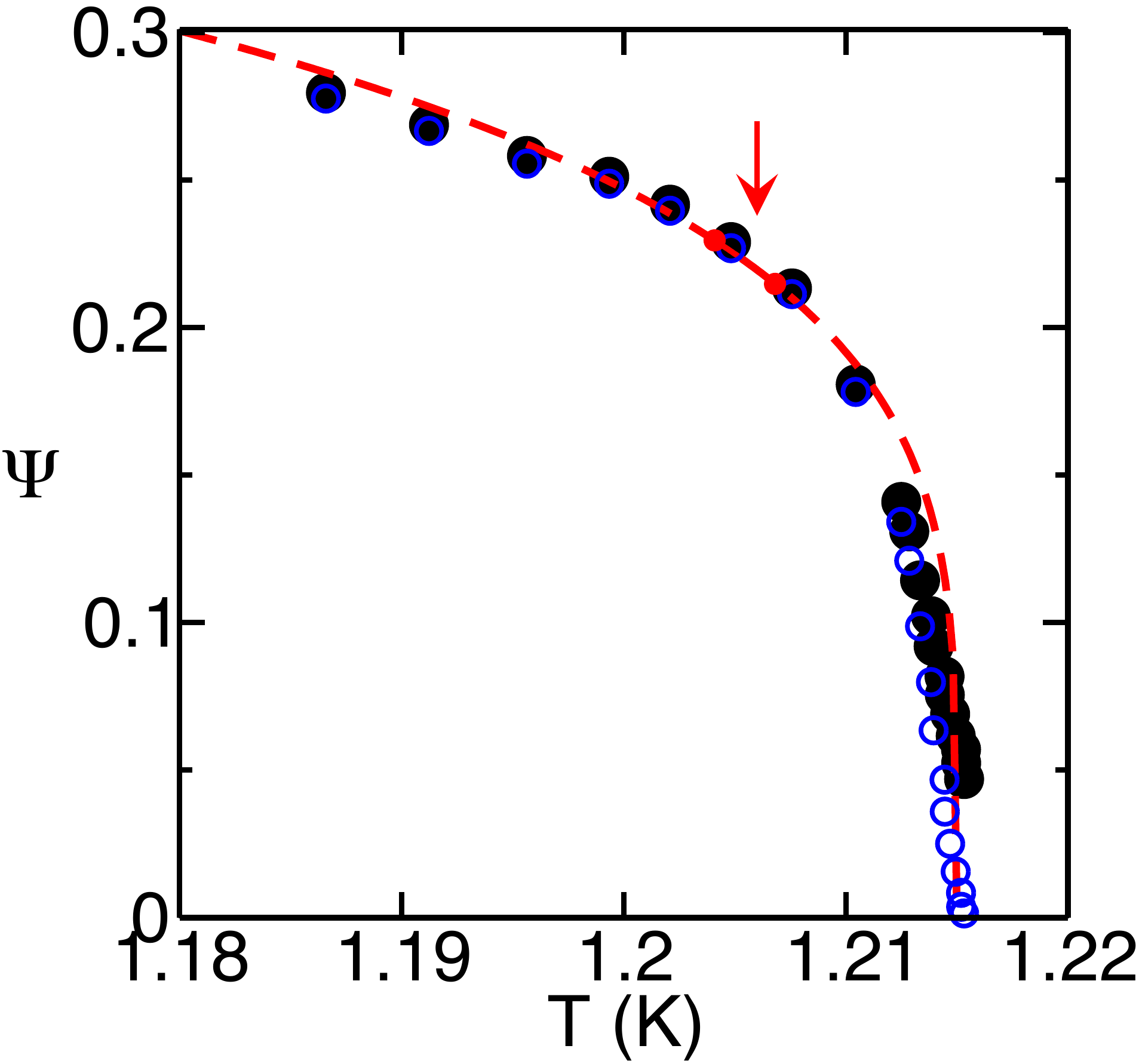}
\caption{Finite-size phase order parameter $\Psi$ (large black filled circles) on length scales of  $>10^{5}$ helium atoms in a superfluid film, derived by transforming the experimental data of Bishop and Reppy~\cite{BR}, using Eq. (\ref{BH}), main text.  
The red line indicates the power law $\Psi = B(T_{\rm C}-T)^{3 \pi^2/128} $ (Eq. (\ref{bh2})), with $B$ and $T_{\rm C}$ calculated from the parameters quoted in Ref.~\cite{BR} -- there is no further fitting in this comparison between theory and experiment. The renormalisation group calculation predicts a close match between theory and experiment only in the range of the full red line (indicated by the arrow) that terminates at $T_{\rm KT}$ and $T^{\ast}$, respectively, in the notation of Ref.~\cite{BH}. However the analytic extension of the power law approximately describes the experimental data over a broader range of temperatures (dotted red line). The small open blue circles indicate data analysed with neglect of dissipation. }   
\label{one}
\end{center}
\end{figure}

\section{Transformation of experimental data}

The above considerations give a method of experimentally estimating the finite-size order parameter in superfluid helium films. 
The integral scale $L$ in the Bishop-Reppy experiment is the dimensionless
dynamical length
\begin{equation}\label{eq:eff_length}
L \simeq \sqrt{\frac{14 \hbar}{m a^2 \omega }},
\end{equation}
where $\omega$ is the measurement frequency and $m$ the mass of a helium atom. 
This $L$ may be interpreted as a diffusion length for vortices, beyond which they couple to the oscillating substrate and destroy the superfluidity.
By fitting the period shift and dissipation data (e.g. Fig. 12 of Ref.~\cite{BR}), Bishop and Reppy determined six parameters~\cite{BR}, of which four relate to the BKT transition:  $L \simeq \exp(12)$, $T_{\rm c} =1.2043$ K, $T_{\rm s} =1.215$ K and $b=5.5$. Here $T_{\rm s}$ is the superfluid transition temperature, $T_{\rm c}$ is the BKT transition temperature and $b$ is a non-universal constant of BKT theory. The parameters $T_{\rm c}$, $T_{\rm s}$ and $b$ are mutually dependent. Redefining $T_{\rm c} \mapsto T_{\rm KT}$, the variables may be transformed to the independent pair of parameters defined in Ref.~\cite{BH}: $T_{\rm s} \mapsto  T_{\rm C}(L) = 1.215 {~\rm K}$,   $b^2/4T_{\rm KT}  \mapsto c(L) = 6.28$ $K^{-1}$.
 
The power law for $\beta$ is entirely determined by these two parameters. 
The specific prediction is~\cite{BH}: 
 \begin{equation}\label{bh2}
\Psi(T) = B(T_{\rm C}-T)^{3\pi^2/128},
\end{equation} 
where $B$ and $T_{\rm C}$ depend on $L$, with 
 \begin{equation}\label{bdef}
B(L)= \left(\frac{1}{\gamma L^2}\right)^{1/16} \left(\frac{3\pi^2}{4 c(L) (\ln L)^2}\right)^{-3\pi^2/128}.
\end{equation} 
The power law (Eq. (\ref{bh2})) is predicted to describe the order parameter only in the vicinity of  a special temperature, $T^{\ast}$, defined such that $\Upsilon(T^{\ast})/kT^{\ast} = 2/\pi$, though in practice it is found to hold over a broader temperature range. It follows from the renormalisation group equations that the two temperatures that characterize the finite-size rounding of the BKT transition are related: $T_{\rm C}(L)-T_{\rm KT}=4(T^{\ast}(L)-T_{\rm KT})=\pi^2/ c(L) (\ln L)^2$.
Using these equations and the experimental values quoted by Bishop and Reppy, we are able to compare the experimental data for helium films with the prediction, Eq. (\ref{bh2}). It should be emphasised that this comparison involves {\it no further free parameters}, the parameters $B$ and $T_{\rm C}$ having been pre-determined as described.

The results are shown in Fig. \ref{one}. Here $\Upsilon(L,T)/kT$ is derived from the reduced period shift $p \equiv 2 \Delta P/P$ and dissipation $q \equiv Q^{-1}$ displayed in Fig. 12 of Ref.~\cite{BR}. The relation is $\Upsilon(L,T)/kT = (2p T_{\rm KT}/C \pi T)(1+ q^2/p^2)$ where $C = 3.4 \times 10^{-6}$ and $T_{\rm KT} = 1.2043$ K~\cite{BR}. This comes from relating $p$ and $q$ (respectively) to the real and imaginary parts of the inverse dielectric function defined in Ref.~\cite{AHNS} and using $\Upsilon/kT = K_{\rm eff}$.
The order parameter $\Psi$ has then been calculated from Eq. (\ref{BH}), while the red lines showing power law behaviour with $\beta = 3\pi^2/128$ have been calculated from Eqs. (\ref{bh2}) and (\ref{bdef}).  

Referring to Fig. \ref{one}, there is excellent agreement between theory and experiment over the predicted range of temperature, where $\Upsilon/kT \gtrsim 2/\pi$ (see Fig. \ref{one}), and  qualitative agreement over a wider range of temperatures, just as for magnetic systems.  Also shown in Fig. \ref{one} is  the data analysed with neglect of dissipation (i.e. approximating $q = 0$). It is confirmed that dissipation is negligible in the temperature range of interest (full red line, Fig. \ref{one}).  

The result of Fig. \ref{one} suggests a method of analysing period shift data in the region where $\Delta P \propto\Upsilon$. Thus, by combining Eqs. (\ref{BH}) and (\ref{bh2}), it follows that
\begin{equation}\label{DP}
\Delta P = \alpha \Upsilon \simeq \frac{- \alpha k T \ln(\gamma L^2)}{8 \pi \left[ \ln B + \beta \ln (T_{\rm C} - T)\right]},
\end{equation}
where $\alpha$ is a scale factor. Reading $T_{\rm C}$ off the $\Delta P$ data as in Ref.~\cite{BR}, and then fitting the data to Eq. (\ref{DP}) by varying $\alpha$ and $B$, allows determination of $\Upsilon = \Delta P/\alpha$ and hence $\Psi$, from Eq. (\ref{BH}). We have tested this method on the experimental data shown in Figs. 3 and 4 of Ref.~\cite{BR}, which cover a sufficient temperature range to allow precise fitting. We find essentially the same result as Fig. \ref{one}, provided that the fitted data is confined to an appropriate range of temperatures somewhat below $T^{\ast}$. Although this method is approximate, it offers a simple practical alternative to the method used to obtain Fig. \ref{one}, which involves the initial step (performed in Ref. \cite{BR}) of fitting the experimental data to the complex numerical solution of the AHNS equations.

\begin{figure*}
\begin{center}
\includegraphics[width=0.95\linewidth]{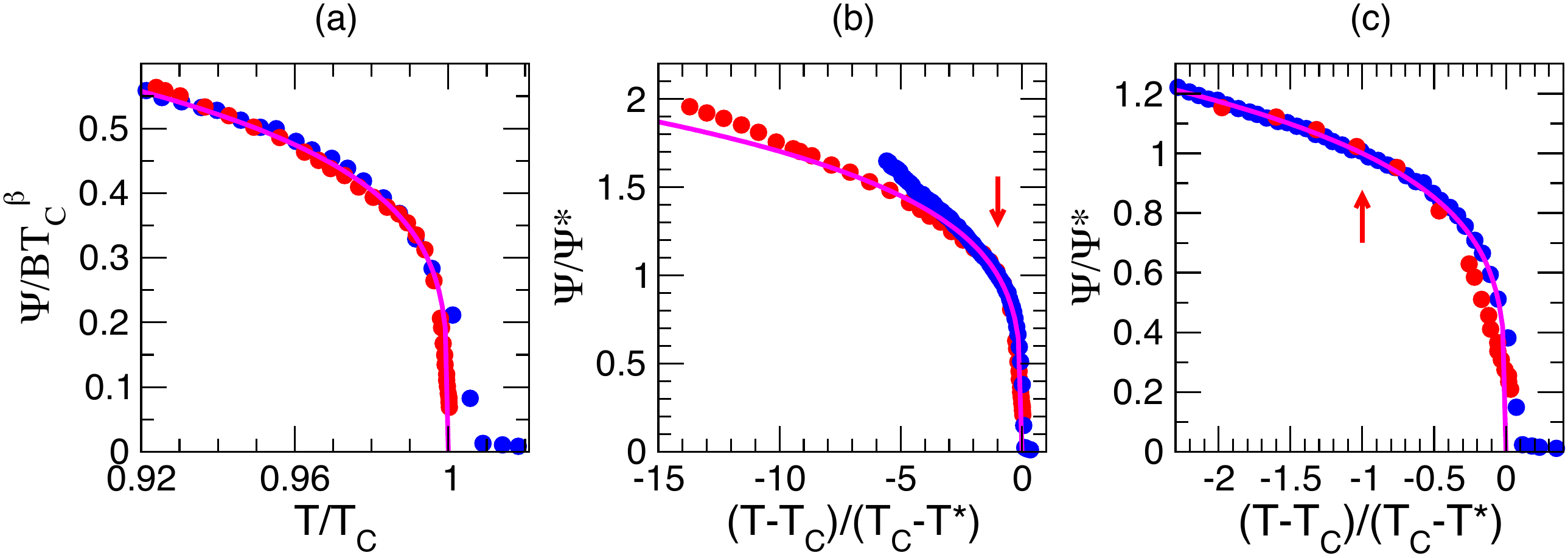}
\caption{Scaling collapse of experimental data sets for the phase order parameter of the helium film of Ref.~\cite{BR} and the magnetic order parameter for a magnetic film of Ref.~\cite{Elmers}. {\bf (a)} Helium data~\cite{BR} ($\Psi/B T_{\rm C}^{3\pi^2/128}$, red points) and scaled magnetic order parameter data~\cite{Elmers} (blue points), collapsed on the temperature scale $T/T_{\rm C}$ using quoted parameters from Refs. \cite{BR, Elmers}. The line is the power law $\Psi/BT_{\rm C}^{3\pi^2/128} = (1-T/T_{\rm C})^{3\pi^2/128}$. Note that the helium data from Fig. 12 of Ref.~\cite{BR} has been extended to lower temperature using data from Fig. 3 of Ref. \cite{BR} analysed with our Eq. (\ref{DP}). 
{\bf (b)}  Reduced helium film phase order parameter  and magnetic thin film order parameters ($\Psi/\Psi^{\ast}$ with $\Psi^{\ast} = (\gamma L^2)^{-1/16}$) {\it vs.} reduced temperature ($(T-T_{\rm C})/(T_{\rm C} - T^{\ast})$). On this normalised, reduced temperature scale, the universal temperature $T^{\ast}$ is at $-1$, as indicated by the red arrow.  Same colour code as in (a).  {\bf (c)}  The same, on an expanded temperature scale in the critical region: theoretically, collapse is expected only near  and below $T^{\ast} =-1$ (red arrow), on the reduced temperature scale.}
\label{three}
\end{center}
\end{figure*}

\section{Universal data collapse}

Until now tests of this universal order parameter scaling have concentrated on numerical simulation and on experimental molecular and magnetic systems, with the latter providing the largest number of data sets \cite{Andrea}. 
The cut-off length scales associated with these experiments ($L \sim 10^2$) are similar to those available from numerics. The present treatment of helium films allows for the extension of this analysis, not only to a wider range of experimental systems but also to 
integral length scales orders of magnitude bigger than those previously studied: $L \sim  10^5$. 

This increased range of experimental finite-size scaling is illustrated in Fig. \ref{three}, where we use Eq. (\ref{bh2}) to collapse together the data for the helium films with magnetisation data  for a  1.6 monolayer sample of iron grown along the cubic (100) direction on a tungsten (100) substrate~\cite{Elmers}, which, as Elmers {\it et al.} elegantly demonstrated, shows 2D-XY behaviour with $L\simeq 140$ through the observed ordering transition. Fig. \ref{three}a shows $\Psi/B T_{\rm C}^\beta$ versus $T/T_{\rm C}$. The data sets fall close to each other and close to the the theoretical prediction developed for the helium film, clearly showing a characteristic form for the order parameter for these diverse systems. However, as the universal exponent, $\beta= 3\pi^2/128$ is predicted at the temperature $T^{\ast}$, which is itself size dependent, a quantitative scaling protocol demands that we plot $\Psi/\Psi^{\ast}$ versus $(T-T_{\rm C})/(T_{\rm C}-T^{\ast})$, which aligns the two universal temperatures at the value $-1$ in the reduced temperature variable (here, $\Psi^{\ast} = (\gamma L^2)^{-1/16}$). The resulting scaling collapse  is shown in Figs. \ref{three}b and \ref{three}c. Excellent quantitative agreement is found between both data sets and our theory at and around $T^{\ast}$, confirming this finite-size scaling analysis as a powerful diagnostic tool for the BKT phase transition for these apparently disparate systems -- the helium film and the magnetic film. Away from $T^{\ast}$, although the data sets remain close to the predicted power law behaviour, some deviation is observed, as indeed the theory predicts.

The reconstruction of the order parameter through Eq. (\ref{BH}) can be tested in detail against direct numerical simulation of XY-type models. We have found that, for easily-simulated system sizes ($L \sim 10^2$), both measures of the order parameter show the predicted scaling behaviour and can be collapsed on to scaling plots, as in Fig. \ref{three}. However, systematic differences appear in the unscaled data which can be attributed to corrections to scaling that we expect to disappear only logarithmically with system size. This difference originates from the $r$ dependence of $K_{\rm eff}(r)$ \cite{Archambault} near $T=T^{\ast}$. We intend to perform a detailed analysis of this behaviour in future work.  

\section{Discussion}

The effective length scale of the helium film (already of order $L\sim 10^5$) will increase as $1/\sqrt{\omega}$, as the measurement frequency goes down and will ultimately be cut off by the true system size if the frequency is driven to zero. It would be interesting to test this prediction experimentally: our Eq. (\ref{DP}) gives a simple theoretical framework with which to do so. Ideally such an experiment would be precisely analogous to that of Bishop and Reppy~\cite{BR}, for as emphasised there, to achieve the BKT transition in helium films depends crucially on the nature of the substrate. 

An appealing aspect of our result is that it associates a universal temperature dependence with experimental data that could previously only be fitted with rather complicated numerical functions~\cite{BR}. It also represents a finite-size scaling approach to helium films that complements others in the literature~\cite{heliumFSS}.
However our theory is presently less complete than that of AHNS in that the condition discussed above, $K_{\rm eff}(r\lesssim L)\simeq K_{\rm eff}(L)$, when extended to $T >  T^{\ast}$, amounts to the neglect of dissipation~\cite{AHNS}. 
Even though dissipation has only a small effect on the derived order parameter (see Fig. \ref{one}), it would be interesting, in the future, to remedy this deficiency: some steps in the right direction were already taken in Ref.~\cite{Archambault}. 

It is tempting to interpret $\Psi$ as a measure of the coherence~\cite{Mayers} of the helium wavefunction. If we accept this, then our analysis illustrates that liquid helium films can show quantum coherence over macroscopic length scales of $l > 10~\mu$m, making them comparable to modern experimental systems such as cold-atom~\cite{Had} and exciton-polaron condensates \cite{Exciton} and coherent over a longer range than electronic quantum transport devices \cite{Webb} or micro-SQUIDS \cite{qbit}. 

Our results strengthen the conclusions of the combined experimental and theoretical works of Bishop and Reppy and of AHNS, showing that they are wholly consistent with finite-size order parameter scaling, which can be used as a key test for the BKT transition~\cite{Wexler}. As a consequence, we are able to propose new experiments and a new protocol for the analysis of experimental data. It would be interesting to relate these results to direct measurements and microscopic calculations of the atomic momentum distribution in helium films~\cite{Diallo}, as well as to measurements of the spin wave stiffness in magnetic systems, both quasi--classical~\cite{BHjump} and quantum~\cite{Ruegg}. 

Our result is also relevant to experiments on two-dimensional Bose-Einstein condensates in cold atomic gases~\cite{Had,Trom}. These are typically confined in optical traps which render them naturally finite sized. In recent years much has been learned about the relation of Bose condensation to superfluidity and the BKT transition through the study of such systems: we refer to Ref.~\cite{Had} for a review. 
Related finite-size effects are indeed broadly relevant to a wide variety of low-dimensional magnetic systems and condensates, including nuclear magnetic films composed of $^3$He~\cite{Casey} and two-dimensional superconductors~\cite{super,super2}. 

In conclusion, while  the remarkable `p-wave' superfluid, helium-3~\cite{He3}, and the highly controllable cold atomic gases~\cite{Had} have largely overtaken it as model systems for studying quantum coherence, ordinary liquid helium II retains a basic fascination as an iconic state of matter -- at once simple and exotic. Our analysis has 
exposed 2D-XY universality in a form that links helium films firmly to magnetic and molecular systems~\cite{Andrea}, that acts over widely varying length scales, and that enables new finite-size scaling experiments. In addition, we have confirmed (Fig. \ref{three}) that magnetic films~\cite{Elmers} afford an equally accurate experimental realisation of the BKT transition as do helium films~\cite{BR}, a fact that has not been widely recognised. We have thus unified alternative approaches to the BKT transition in superfluids~\cite{BR} and magnets~\cite{Als,Willis, Elmers, Hjorvarsson, Andrea}, and confirmed the universal application of BKT theory~\cite{B,KT,K,JKKN,NK,AHNS} to experimental systems.





\acknowledgments
It is a pleasure to thank Tommaso Roscilde for useful discussions concerning this work.  
M.F.F. is grateful for financial support from the CNRS,  University  College London and ANR JCJC-2013 `ArtiQ'. P.C.W.H. acknowledges financial  support from the Institut Universitaire de France.




\end{document}